\documentclass[aps,longbibliography,
  twocolumn,amsmath,
  amssymb,nofootinbib, superscriptaddress, preprintnumbers]{revtex4-1}

\usepackage{graphics}      
\usepackage{graphicx}      
\usepackage{longtable}     
\usepackage{url}           
\usepackage{bm}            
\usepackage{xcolor}
\usepackage{mathrsfs}
\usepackage[colorlinks,allcolors=blue]{hyperref}
\usepackage{blindtext}
\usepackage{hyperref}
\usepackage{float}
\usepackage{bbold}
\usepackage{lipsum}
\usepackage[normalem]{ulem}
\usepackage{orcidlink}
\usepackage{academicons}
\usepackage{xcolor}
\usepackage{subfigure}
\usepackage{booktabs}
\usepackage{hyperref}

\newcommand{\orcid}[1]{\href{https://orcid.org/#1}{\textcolor[HTML]{A6CE39}{\aiOrcid}}}

\newcommand{\sa}[1]{\textcolor{black}{#1}}

\newcommand{\mrw}[1]{{\color{magenta}{#1}}}

\begin{document}
\preprint{MPP-2023-270}

\title{Physics-Informed Neural Networks for Predicting the Asymptotic Outcome of Fast Neutrino Flavor Conversions }

\newcommand*{\MPP}{\textit{\small{Max-Planck-Institut f\"ur Physik (Werner-Heisenberg-Institut), Boltzmannstr. 8, 85748 Garching, Germany}}}

\author{Sajad Abbar \orcidlink{0000-0001-8276-997X}   } 
\affiliation{\MPP}
\author{Meng-Ru Wu \orcidlink{0000-0003-4960-8706}}
\affiliation{Institute of Physics, Academia Sinica, Taipei, 11529, Taiwan}
\affiliation{Institute of Astronomy and Astrophysics, Academia Sinica, Taipei, 10617, Taiwan}
\affiliation{Physics Division, National Center for Theoretical Sciences, Taipei 10617, Taiwan}

\author{Zewei Xiong \orcidlink{0000-0002-2385-6771} }
\affiliation{GSI Helmholtzzentrum {f\"ur} Schwerionenforschung, Planckstra{\ss}e 1, D-64291 Darmstadt, Germany}

\begin{abstract}

In the most extreme astrophysical environments, such as core-collapse supernovae (CCSNe) and neutron star mergers (NSMs), neutrinos can undergo fast flavor conversions (FFCs) on exceedingly short scales. Intensive  simulations  have demonstrated that FFCs can attain equilibrium states in certain models. 
In this study, we utilize physics-informed neural networks (PINNs) to \sa{predict} the asymptotic outcomes of FFCs, 
 by specifically targeting the first two moments of neutrino angular distributions. 
 This \sa{makes} our approach suitable for state-of-the-art CCSN and NSM simulations.
Through effective feature engineering and the incorporation of customized loss functions that penalize discrepancies in the predicted total number of $\nu_e$ and $\bar\nu_e$, our PINNs demonstrate remarkable accuracies, with an error margin of  $\lesssim3\%$.
Our study represents a substantial leap forward in the potential incorporation of FFCs into  simulations of CCSNe and NSMs, thereby \sa{enhancing} our \sa{understanding} of these extraordinary astrophysical events.

 \end{abstract}

\maketitle

\section{Introduction}

Core-collapse supernovae (CCSNe) and neutron star mergers (NSMs) \sa{are among the most extreme astrophysical phenomena}, 
representing the \sa{death}  of massive stars' life cycles and the collision of \sa{extremely} dense remnants, respectively. These \sa{events} not only signify the demise of massive stars and dense objects but also reveal some of the universe's most energetic and mysterious phenomena.

At the heart of these \sa{extraordinary settings} lies a captivating process: the emission of neutrinos, released in copious amounts during both CCSNe and NSMs~\cite{Burrows:2020qrp, Janka:2012wk, Foucart:2022bth, Kyutoku:2021icp, Colgate:1966ax, Lattimer:1974slx}. During their propagation through the  extreme conditions within these events, neutrinos undergo a fascinating phenomenon known as \emph{collective} neutrino oscillations~\cite{pantaleone:1992eq, sigl1993general, Pastor:2002we,duan:2006an, duan:2006jv, duan:2010bg, Mirizzi:2015eza}~(for a recent review see Ref.~\cite{Volpe:2023met}).
This intriguing behavior emerges from the intricate interplay between the propagating neutrinos and 
  the dense background neutrino gas, where coherent forward scatterings play a crucial role. This phenomenon behaves in a nonlinear and collective manner, leading to a complex tapestry of flavor conversions.

Of particular interest are the so-called \emph{fast} flavor conversions (FFCs), which occur on scales characterized by $\sim (G_{\rm{F}} n_{\nu})^{-1}$ 
(see, e.g., Refs.~\cite{Sawyer:2005jk, Sawyer:2015dsa,
Chakraborty:2016lct, Izaguirre:2016gsx,Capozzi:2017gqd, Wu:2017qpc, Wu:2017drk, Abbar:2017pkh, Abbar:2018beu,Capozzi:2018clo, 
Martin:2019gxb, Abbar:2018shq, Abbar:2019zoq, Capozzi:2019lso, George:2020veu, Johns:2019izj, Martin:2021xyl, Tamborra:2020cul,  Sigl:2021tmj,   Morinaga:2021vmc, Nagakura:2021hyb,  Sasaki:2021zld, Padilla-Gay:2021haz, 
Xiong:2021dex, Capozzi:2020kge, Abbar:2020qpi, Capozzi:2020syn, DelfanAzari:2019epo, Harada:2021ata,  Abbar:2021lmm, Just:2022flt, 
Padilla-Gay:2022wck, Capozzi:2022dtr, Zaizen:2022cik, Shalgar:2022rjj,  Kato:2022vsu, Zaizen:2022cik,  Bhattacharyya:2020jpj, Wu:2021uvt, Richers:2021nbx, Richers:2021xtf, Dasgupta:2021gfs, Capozzi:2022dtr, Nagakura:2022kic, Ehring:2023lcd, Ehring:2023abs, Xiong:2023upa, Fiorillo:2023hlk, Nagakura:2023wbf, Martin:2023gbo, Fiorillo:2023mze,
Grohs:2023pgq}).
 Here, $G_{\rm{F}}$ represents the Fermi coupling constant, and $n_{\nu}$ denotes the neutrino number density. These FFCs can take place on scales much shorter than what would be expected in the vacuum.
FFCs occur \emph{iff} the angular distribution of the neutrino lepton
number,  defined as,
\begin{equation}
\begin{split}
  G(\mathbf{v}) =
  \sqrt2 G_{\mathrm{F}}
  \int_0^\infty  \frac{E_\nu^2 \mathrm{d} E_\nu}{(2\pi)^3}
        &[\big( f_{\nu_e}(\mathbf{p}) -  f_{\nu_x}(\mathbf{p}) \big)\\
              &- \big( f_{\bar\nu_e}(\mathbf{p}) -  f_{\bar\nu_x}(\mathbf{p}) \big)],
 \label{Eq:G}
\end{split}
\end{equation}
crosses zero at some $\mathbf{v} = \mathbf{v}(\mu,\phi_\nu)$, with $\mu =\cos\theta_\nu$~\cite{Morinaga:2021vmc}. 
Here, $E_\nu$, $\theta_\nu$, and $\phi_\nu$ are the neutrino energy,  
the zenith, and azimuthal angles of the neutrino velocity, respectively. 
The $f_{\nu}$'s are the neutrino 
occupation numbers of different flavors, with $\nu_x$ and $\bar\nu_x$ denoting the heavy-lepton flavor of neutrinos and antineutrinos. 
When  $\nu_x$ and $\bar\nu_x$ have similar angular distributions, a scenario commonly observed in state-of-the-art 
CCSN simulations, this expression transforms into the conventional definition of the neutrino electron lepton number, $\nu$ELN.

FFCs tend to occur on spatial and temporal scales which are expected to be significantly shorter than those typically addressed in hydrodynamical simulations of CCSNe and NSMs. As a consequence, integrating FFC into these simulations presents a notable challenge. A potential strategy to \sa{address} this challenge involves breaking down the problem into two scale hierarchies. 
This consideration motivates 
conducting local dynamical simulations at shorter scales and subsequently \sa{integrating} the findings into some practical prescriptions. \sa{Such} prescriptions can then be efficiently applied to the broader astrophysical modelings and hydrodynamic simulations~\cite{xiong2020potential,George:2020veu,Li2021a,Just:2022flt,Fernandez:2022yyv,Ehring:2023abs, Ehring:2023lcd}.

The assessment of the outcome of FFCs has undergone thorough examinations of local dynamical simulations conducted within confined spaces employing periodic boundary conditions~\cite{Bhattacharyya:2020dhu,Bhattacharyya:2020jpj,Wu:2021uvt,Richers:2021nbx,Zaizen:2021wwl,Richers:2021xtf,Bhattacharyya:2022eed,Grohs:2022fyq,Abbar:2021lmm,Richers:2022bkd,Zaizen:2022cik,Xiong:2023vcm} (see also Ref.~\cite{Zaizen:2023ihz} for the possible impact of the choice of boundary conditions). Insights from these investigations indicate a tendency toward kinematic decoherence in flavor conversions, generally \sa{resulting in} quasistationary states. 
These stationary states can be characterized by survival probabilities, which are governed by the conservation of neutrino lepton number and have been demonstrated to be possibly modeled by analytical formulation to a good accuracy~\cite{Xiong:2023vcm}.

Despite the  existence of such analytical formulae  for the angular distribution of survival probabilities, implementing FFCs  in CCSN and NSM simulations remains a challenge. The obstacle lies in the requirement of having access to complete angular distributions of neutrinos to determine FFC outcomes through these analytical expressions. However, acquiring such detailed angular information proves challenging in most cutting-edge CCSN and NSM simulations due to their computationally intensive nature.

As a \sa{practical} alternative to considering the full neutrino angular distributions, many state-of-the-art simulations opt for a more feasible approach by simplifying neutrino transport through a limited set of angular distribution moments~\cite{Shibata:2011kx, Cardall:2012at, thorne1981relativistic}. In our specific investigation, we concentrate on radial moments, defined as,
\begin{equation}
I_n = \int_{-1}^{1} \mathrm{d}\mu\ \mu^n\ \int_0^\infty \int_0^{2\pi} \frac{E_\nu^2 \mathrm{d} E_\nu \mathrm{d} \phi_\nu}{(2\pi)^3} \
        f_{\nu}(\mathbf{p}).
\end{equation}
These moments effectively capture crucial aspects of the neutrino angular distribution while facilitating a computationally more manageable treatment\footnote{Our primary focus is presently on axisymmetric crossings, particularly emphasizing radial moments where the angular distribution integrates over  $\phi_\nu$. It's crucial to highlight that our current study excludes non-axisymmetric crossings. Exploring these aspects is a subject reserved for future investigations.}.


In practical scenarios, one often encounters a situation where simulations directly provide only the first two moments, $I_0$ and $I_1$, for (anti)neutrinos. 
\sa{The problem then becomes determining}
the ultimate values of $I_0$ and $I_1$ following FFCs, based on their initial states. \sa{Note that} despite the availability of analytical formulae for the angular distribution of the neutrino survival probabilities, determining these final values is inherently complex.

This paper represents a pioneering effort in predicting the asymptotic outcomes of FFCs  in the moments scenario, using artificial neural network (NNs).
\sa{NNs imitate closely the brain's network of connected neurons. Specifically, they have layers of artificial neurons that handle information. NNs have been proven to be useful in solving tricky problems due their strong learning capacities, resulting from 
adjusting the connections between neurons during the training phase. Their ability to learn from data without explicit
programming sets NNs apart, 
making them attracting tools
across various domains.}
In particular, NNs have been extensively used in  the field of astrophysics and high energy physics~\cite{Guest:2018yhq, smith2023astronomia}.

Our approach involves the utilization of a NN, which takes 
the essential information \sa{extracted from the initial} (anti)neutrino zeroth and first moments and then outputs the  corresponding moments regarding the asymptotic outcome of  FFCs. 
In particular, we employ physics-informed neural networks (PINNs), where the learning and  performance of the NN   can be enhanced with the utilization of the domain knowledge (specialized information specific to the problem that can be integrated into the  NN)~\cite{karniadakis2021physics,cuomo2022scientific, raissi2019physics}.
Our findings demonstrate the efficacy of a single hidden layer PINN, achieving a
remarkable accuracy 
for the asymptotic values of $I_0$ and $I_1$.

The paper is structured as follows. 
 In Sec.~\ref{sec:FFC}, we initiate by detailing our simulations of FFCs and elucidating the assumptions  to deriving the outcomes of FFCs. Moving forward in Sec.~\ref{sec:PINN}, we delve into the architecture of our NNs, shedding light on the requisite feature engineering and the deployment of customized loss functions. The ensuing discussion encompasses the results gleaned from both two- and three-flavor scenarios. Finally, we conclude in Sec.~\ref{sec:dis}.

  \section{FFCs simulations}\label{sec:FFC}
  
 To effectively train our NN, we require a substantial number of training samples containing initial values of the (anti)neutrino moments, $I_0$ and $I_1$, within the neutrino gas. These samples should also encompass their corresponding final values, reflecting the asymptotic outcomes of FFCs.

Our chosen physical model involves the evolution of FFCs within a one-dimensional (1D) box, mirroring the setup outlined in Ref.~\cite{Wu:2021uvt}. This model assumes translation symmetry along the $x$ and $y$ axes, axial symmetry around the $z$ axis, and periodic boundary conditions in the $z$ direction. Notably, we omit considerations of vacuum mixing and neutrino-matter forward scattering in this model.

In this study, we prime the neutrino gas for tracking its flavor evolution by employing two widely utilized parametric neutrino angular distributions documented in existing literature.
The first one is the maximum entropy distribution
defined as, 
\begin{equation}
f^{\rm{max-ent}}_\nu(\mu) = \exp[\eta + a\mu],
\end{equation}
where we here consider the $\phi_\nu$-integrated distribution, i.e.,  
\begin{equation}
 f_{\nu}(\mu) =  \int_0^\infty \int_0^{2\pi} \frac{E_\nu^2 \mathrm{d} E_\nu \mathrm{d} \phi_\nu}{(2\pi)^3} 
        f_{\nu}(\mathbf{p}).
\end{equation}
This is a very natural choice for the neutrino angular distribution since
the maximum entropy closure \cite{Cernohorsky:1994yg} is currently very popular  in the  moment-based neutrino
transport methods. 
This  parametric  distribution has been also used to detect $\nu$ELN crossings using fitting and machine learning techniques~\cite{Richers:2022dqa, abbar2023applications,Abbar:2023zkm}.
Another angular distribution considered in the literature of FFCs (see, e.g., Refs.~\cite{Wu:2021uvt, Yi:2019hrp})  is the Gaussian distribution defined as,
\begin{equation}
f^{\rm{Gauss}}_\nu(\mu) = A\exp[-\frac{(1-\mu)^2}{\xi}].
\end{equation}
Note that both of these distributions have a parameter which determines the overall neutrino number density, namely
$\eta$ and $A$, and the other parameters  determining the shape of the distribution, i.e., $a$ and $\xi$.
Allowing for two distinct forms of angular distributions takes into consideration potential deviations in the shape of neutrino angular distributions in realistic simulations, which can occur, e.g., due to the use of different closure relations.

To ready our datasets, we begin with the initial angular distributions of neutrinos, which can either follow a maximum entropy distribution or a Gaussian distribution. Subsequently, we utilize analytical neutrino survival probabilities to determine the asymptotic outcome of FFCs. By performing integration over the neutrino angular distributions, we can obtain the initial and final values of $I_0$ and $I_1$.

In our analytical treatment of the survival probability, we follow closely our recent work in Ref.~\cite{Xiong:2023vcm}. 
We assume that 
 $G(\mu)\ (=\int_{0}^{2\pi}\mathrm{d}\phi_\nu G(\mathbf{v}))$ has only one zero crossing $\mu_c$.
 This helps us to define,
 \begin{equation} 
 \begin{split}
 \Gamma_{+}&=\bigg| \int_{-1}^{1} \mathrm{d}\mu\ G(\mu) \Theta[G(\mu)] \bigg|, \\
 \Gamma_{-}&=\bigg| \int_{-1}^{1} \mathrm{d}\mu\ G(\mu) \Theta[-G(\mu)] \bigg|,
 \end{split}
 \end{equation} 
 as the integration of positive and negative parts of $G(\mu)$.
Here $\Theta$ is the Heaviside theta function. In the following, we specify the $\mu$ range
over which the above integral is smaller (larger) by $\mu^{<} (\mu^{>})$. For the survival probability in the two-flavor scenario,
we  use the analytical formula:

\begin{equation}\label{eq:sur}
    P^{\mathrm{2f}}_{\mathrm{sur}}(\mu) = 
    \begin{cases}
        \frac{1}{2} & {\rm for~}\mu^<, \\
        \mathcal{S}(\mu) & {\rm for~}\mu^>,
    \end{cases}
\end{equation}
where the distribution over $\mu^>$ is formulated as,
\begin{equation}\label{eq:express_continuous}
    \mathcal{S}(\mu) = 1-\frac{1}{2}h(|\mu-\mu_c|/\zeta).
\end{equation}
Here,  $h(x)$ is a $\mu$-dependent function that monotonically decreases from 1 to 0 when $x$ increases from 0 to infinity.
To be specific, we here assume $h(x)$ to have  a power-1/2 form, i.e.,  $h(x) = (x^2+1)^{-1/2}$. In addition, the parameter $\zeta$ can be found such that the survival
probability function is continuous.

In the three-flavor case where $\nu_\mu$ and $\nu_\tau$ are indistinguishable, 
one can simply find the survival probabilities by 
the expression  $P^{\mathrm{3f}}_{\mathrm{sur}}(\mu) = 1-4[1-P^{\mathrm{2f}}_{\mathrm{sur}}(\mu)]/3$.

 As illustrated in Table.~1 of Ref.~\cite{Xiong:2023vcm}, adopting a power-1/2 form for the survival probability proves to yield a comparatively low error in computing $I_0$ and $I_1$ analytically. This \sa{explains} the rationale behind opting for this analytical survival probability in this work. In the concluding part of Sec.~\ref{sec:PINN}, we \sa{also explore} the scenario where the outcomes derived from actual simulations of FFCs  are applied.

\section{Applications of Neural Networks}\label{sec:PINN}

Before unveiling our findings, it's crucial to emphasize 
that to ensure  high performance of our NN models in the test set, it is necessary to divide the dataset into three distinct sets.
These sets are defined as follows: i) Training Set: This set serves as the foundation for training the NN, allowing it to learn and adapt based on the provided data. ii) Development Set: Also known as the validation set, this subset plays a pivotal role in determining the optimal hyper-parameters of the algorithm. It serves as a testing ground to fine-tune the model for optimal performance. iii) Test Set: To assess the NN efficacy on novel, unseen data, the test set is utilized. This set provides a critical evaluation of the model's generalization capabilities beyond the training data.

\subsection{The architecture of NNs}

For a given arbitrary neutrino gas, one is provided with the initial values of $I_0$'s and $I_1$'s of $\nu_e$, $\bar\nu_e$, and $\nu_x$. In this context, we make the assumption that the initial distributions of $\bar\nu_x$ and $\nu_x$ are identical (though their final ones following FFCs could be different), a simplification that aligns with the majority of state-of-the-art CCSN and NSM simulations.
In order to enhance the performance of our NNs, we introduce a layer of feature engineering, employing the following features as pertinent inputs in our NNs:

\begin{equation}\label{eq:inputs}
\begin{split}
&\alpha,\quad \alpha_{\nu_x},\quad F_{\nu_e},\quad F_{\bar\nu_e},\quad \mathrm{and}\quad F_{\nu_x},\\
\mathrm{with:}\quad  &\alpha = \frac{n_{\bar\nu_e}}{n_{\nu_e}},\quad \alpha_{\nu_x} = \frac{n_{\nu_x}}{n_{\nu_e}}, \quad \mathrm{and}\quad F_{\nu} = \bigg( \frac{I_1}{I_0} \bigg)_{\nu}.
\end{split}
\end{equation}

Note that the selection of these features offers explicit insights into the configuration of neutrino angular distributions, which \sa{plays a crucial role} in understanding the asymptotic outcome of FFCs. 
Moreover, with the provided values of $I_0$ and $I_1$ for a specific neutrino species (derivable from $\alpha_{(\nu_x)}$ and $F_\nu$), 
we use the root-finding function \emph{fsolve} in \textsc{Python} 
to determine  the complete shape of the neutrino angular distribution, being either maximum entropy or the Gaussian one.
Furthermore, it is worth highlighting that all quantities in this context are normalized by the initial $\nu_e$ number density, allowing the convenient choice of setting it to $n^{\rm{initial}}_{\nu_e} = 1$. This simplification  reduces the number of inputs to our NNs, and notably, there is no input parameter related to $n_{\nu_e}$.


Though the aforementioned features serve as a necessary foundation for developing a NN, there remains room for further enhancement through \sa{more advanced} feature engineering to optimize the performance of our NNs. This optimization can be achieved by gaining insights from the neutrino survival probability's shape, as expressed in Eq.~(\ref{eq:sur}). Substantial information pertaining to the distribution of the survival probability can be derived by learning the position of $\mu_c$. Another 
valuable piece of information, given $\mu_c$, is determining the specific side of $\mu_c$ on which equipartition occurs, 
while the behavior of the survival probability on the other side is governed by conservation laws.
Information regarding the side on which equipartition occurs is provided in the quantity
 $E_{RL}$, a binary number which is 1 if the equipartition occurs for $\mu_c\leq\mu$,
and 0 otherwise.

\begin{figure} [tb]
\centering
\begin{center}
\includegraphics*[width=.4\textwidth, trim= 0 1 0 0, clip]{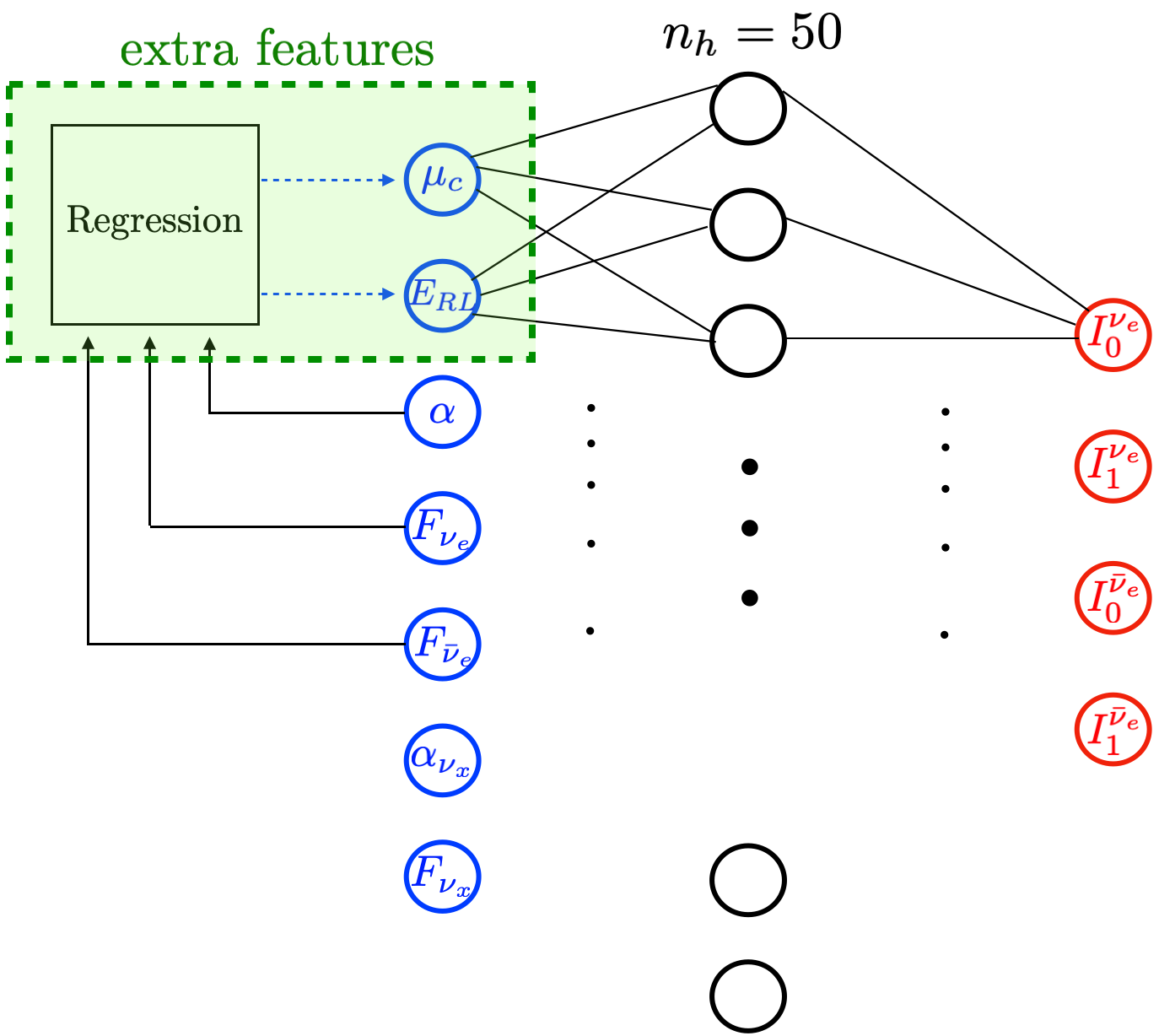}
\end{center}
\caption{
Schematic architecture of our NNs. The green zone shows the implementation of the extra features, $\mu_c$,
and $E_{RL}$, which are obtained through an extra layer of regression, using  linear and logistic regressions, respectively. Here, $\mu_c$ is the
crossing direction and $E_{RL}$ is a binary, which is 1 if the equilibrium occurs for $\mu_c\leq\mu$,
and 0 otherwise. In our basic NN, referred to as the NN with no extra features, the NN only takes the inputs highlighted in Eq.~(\ref{eq:inputs}).
However and in our PINN, we provide our NN with the extra features $\mu_c$  and $E_{RL}$.
}
\label{fig:scheme}
\end{figure}

In our NN framework, we explore two distinct architectures, as illustrated in Fig.~\ref{fig:scheme}. In the foundational architecture, we integrate only $\alpha$, $\alpha_{\nu_x}$, $F_{\nu_e}$, $F_{\bar\nu_e}$, and $F_{\nu_x}$ into our NN. An alternative NN that we examine involves an additional layer of feature engineering, as discussed in the preceding paragraph, encompassing information about $\mu_c$ and $E_{RL}$.

Practically, this augmentation is accomplished by constructing a \emph{separate} regression model trained on our dataset, from which information regarding $\mu_c$ and $E_{RL}$ can be readily extracted. \sa{We have confirmed}  that 
the computation of $E_{RL}$ and $\mu_c$ can be performed with  small errors.

As illustrated  in Fig.~\ref{fig:scheme}, our feedforward NN has a single hidden layer containing 50 neurons, unless stated otherwise. The rationale for this choice  is illustrated in Fig.~\ref{fig:num_hid} and the text around it.
 Also regarding the output layer, our NNs provide the values of $I_0$ and $I_1$ for both $\nu_e$ and $\bar\nu_e$, effectively utilizing a total of 4 neurons. The determination of $I_0$ and $I_1$ for $\nu_x$ and $\bar\nu_x$ can be deduced by applying the principles of neutrino and antineutrino number density and momentum conservation.
 In simpler terms, our NN's ensures that the fundamental laws governing neutrino conservation are \sa{respected} without
 any exceptions.

 \subsection{Loss functions}

The loss function, $\mathcal{L}$, is a crucial component in training NNs, serving as a measure of the model's predictive performance. It quantifies the disparity between predicted values and actual target values, providing a guide for the model to adjust its parameters during the optimization process.

When it comes to neutrino flavor conversions in CCSNe and NSMs, a critical parameter of utmost significance is the number of neutrinos in the electron channel, i.e., $ N_{{\nu_e} + {\bar\nu_e}} = n_{\nu_e} + n_{\bar\nu_e}$, as opposed to the number of neutrinos in the heavy-lepton channel, $ N_{{\nu_x} + {\bar\nu_x}} = n_{\nu_x} + n_{\bar\nu_x}$. Leveraging this crucial physical insight to enhance the performance of our NN, we incorporate an additional loss term in the optimization of the NN model with the extra features. This  loss term is 
 designed to penalize discrepancies in $N_{{\nu_e} + {\bar\nu_e}}$, and is defined as,
 \begin{equation}
   \mathcal{L}_{\rm{extra}} = \frac{1}{N_{\rm{sample}}} \Sigma_i (\Delta N_{{\nu_e} + {\bar\nu_e}, i})^2,  
 \end{equation}
  where  $\Delta$, $N_{\rm{sample}}$, and $\Sigma_i$ denote the difference between the true and predicted values, the number of samples in the training set, and the summation over the training samples, respectively. This  loss term
 provides an additional constraint for the model.
The integration of the domain knowledge characterizes this particular NN architecture as a PINN, given that its distinctive nature is shaped by our insights into the underlying physics of the problem~\cite{karniadakis2021physics,cuomo2022scientific, raissi2019physics}. 
 The PINN should be compared with our basic NN, referred to as NN with \emph{no extra features}, for which the loss term only includes the ordinary mean squared errors of the output parameters.
 Note that the incorporation of the domain knowledge in our PINN includes both
 architectural aspects, utilizing additional features, and learning-based enhancements via the loss function.
 It's also important to highlight that unlike what typically observed in PINNs, our approach does not involve a loss term associated with some partial differential equations.

 \subsection{Three-flavor scenario}

\begin{figure*} [tb]
\centering
\begin{center}
\includegraphics*[width=1.\textwidth, trim= 90 40 100 60, clip]{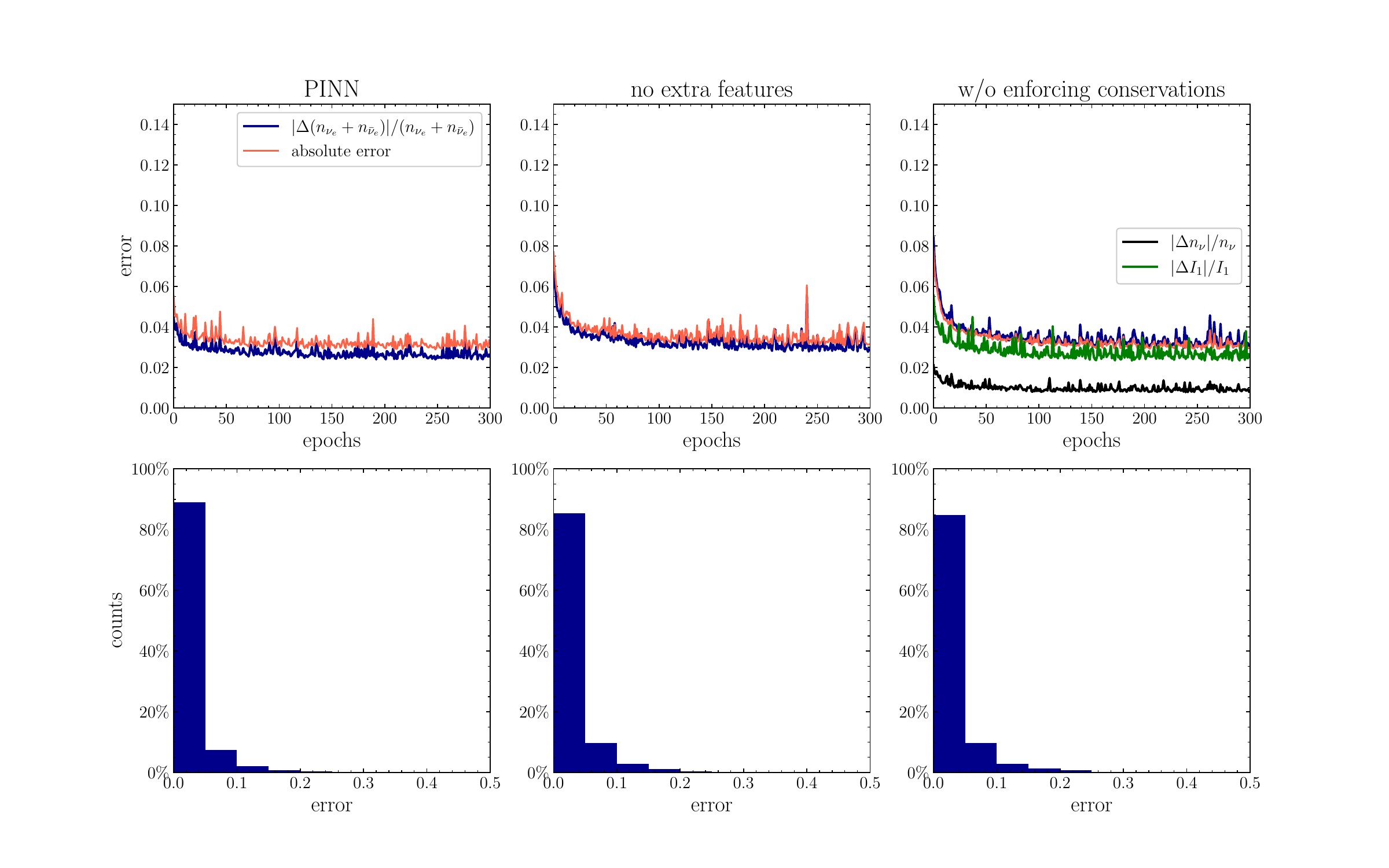}
\end{center}
\caption{
Performance evaluation of various NN models.
In the upper panels, we present the absolute error in the output parameters of the NN models, along with the relative error in the total number of neutrinos within the electron channel,  $N_{{\nu_e} + {\bar\nu_e}}$. Moving to the lower panels, one can find the distribution in the error  of $N_{{\nu_e} + {\bar\nu_e}}$. The results are showcased for our PINN, our basic NN with no extra features, as well as a NN model that doesn't enforce the conservation laws for neutrinos, respectively. In the upper right panel, the black and green curves represent the relative error in the total number and its first moments for (anti)neutrinos.
}
\label{fig:error}
\end{figure*}

In this section, we \sa{discuss  the evaluation of our NNs concerning their predictions for the asymptotic outcome of FFCs  in the three-flavor scenario.}
To train and assess our model, we utilize a dataset comprising a well-balanced combination of maximum entropy and Gaussian initial neutrino angular distributions. The total size of our dataset is $2\times 10^5$ points.
The ultimate outcome of FFCs is determined through a three-flavor survival probability, as detailed in Eq.~(\ref{eq:sur}) and the surrounding text. To better encapsulate realistic conditions regarding the values of $n_\nu$'s and the hierarchy among $F$'s, we  prepare each sample by randomly selecting the inputs for our NNs. 
Specifically, we set $\alpha \in (0,2.5)$,  $\alpha_{\nu_x} \in (0,3)$, $F_{\nu_x} \in (0,1)$,  $F_{\bar\nu_e} \in (0.4F_{\nu_x}, F_{\nu_x})$, and $F_{\nu_e} \in (0.4 F_{\bar\nu_e}, F_{\bar\nu_e})$.
This selection process ensures consistency with the expected hierarchy $F_{\nu_e} \lesssim F_{\bar\nu_e} \lesssim F_{\nu_x}$, characteristic of CCSN environment. Given these quantities, one can then determine the initial angular distributions of neutrinos. Using the analytical survival probability, the final $I$'s can be derived.


In the left panels of Fig.~\ref{fig:error}, we present the performance results of our PINN model. 
Here, an epoch  refers to a single pass through the entire training dataset during the training phase. 
Notably, the relative error in the electron neutrino number density, defined as $|\Delta (n_{\nu_e} + n_{\bar\nu_e})|/(n_{\nu_e} + n_{\bar\nu_e})$, can attain values as low as 2.5\%. Furthermore, we observe that the mean  absolute 
error 
in the output quantities, defined as  $(|\Delta I^{\nu_e}_0| + |\Delta I^{\nu_e}_1| + |\Delta I^{\bar\nu_e}_0| + |\Delta I^{\bar\nu_e}_1|)/4$,  can reach values $\sim 3\%$. Additionally, from the lower panel, it is evident that almost 90\% of the predictions exhibit errors of  $\lesssim 5\%$ in $N_{{\nu_e} + {\bar\nu_e}}$.


In the middle panels of Fig.~\ref{fig:error}, we present the performance of our basic NN with no extra feature and 
no loss term for enhancing the accuracy of $N_{{\nu_e} + {\bar\nu_e}}$.
It is evident that the errors in this case are a bit greater than those of the PINN model.
It's also worth noting that while the error in
$N_{{\nu_e} + {\bar\nu_e}}$ is smaller than the absolute error in the output quantities, the gap between them has been  reduced. 
This could be attributed to the lack of a specific loss term targeting the reduction of errors in $N_{{\nu_e} + {\bar\nu_e}}$. 
 In addition and by examining the lower panel, we can observe that 
almost 85\% of the predictions still exhibit errors of $\lesssim5\%$ in $N_{{\nu_e} + {\bar\nu_e}}$.

\begin{figure*} [tb]
\centering
\begin{center}
\includegraphics*[width=1.\textwidth, trim= 95 0 100 20, clip]{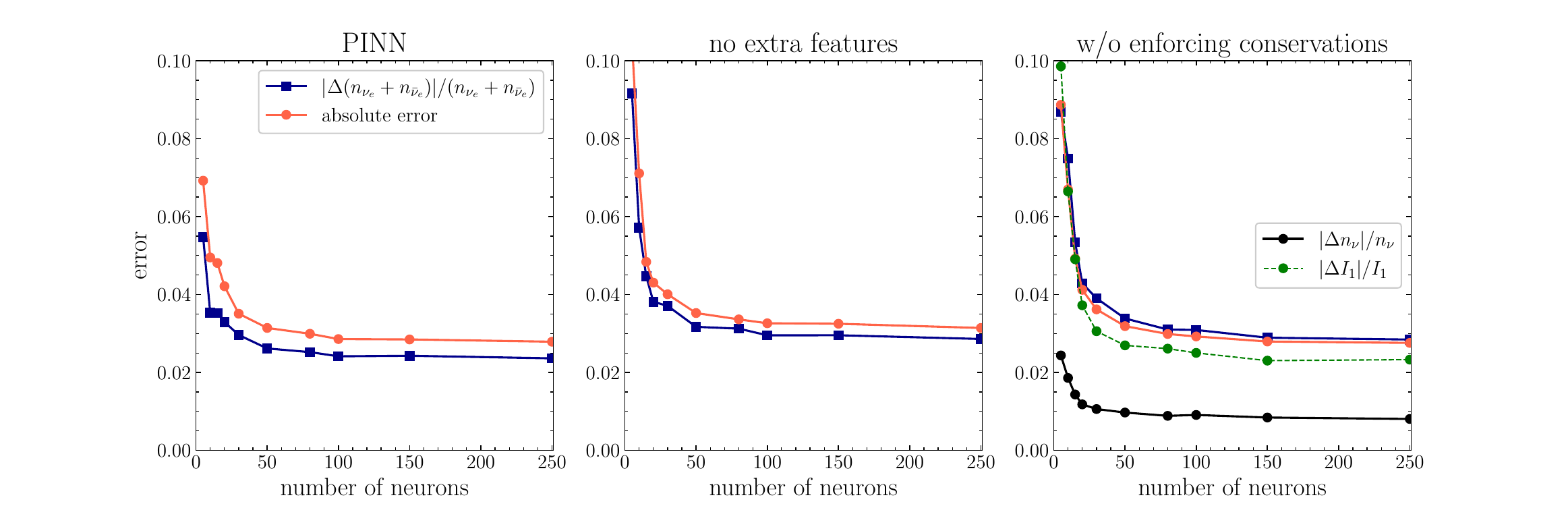}
\end{center}
\caption{
Performance evaluation of various NN models (on the validation set) as a function of the number of neurons in the hidden layer.
It is evident that the NNs perform best on the validation set once $n_h\gtrsim 50$.
The labels and NN models are the same as those in Fig.~\ref{fig:error}.
}
\label{fig:num_hid}
\end{figure*}

Instead of rigidly enforcing the conservation of neutrino quantities, an alternative approach \sa{involves a more flexible NN model for which strict conservation laws are not forcefully respected. } 
Instead, one can aim to derive all the neutrino moments ($I_0$'s and $I_1$'s for all flavors) as outputs, while introducing an additional loss term, which \emph{effectively} enforces the conservation of (anti)neutrino number densities and momenta.
This increased flexibility, combined with the loss term addressing conservation, has the potential to enhance the training of the NN,  resulting in reduced errors in its predictions.
Consider that this NN architecture could also be referred to as a PINN due to its loss term encompassing domain-specific knowledge,  accounting for the conservation laws governing neutrino number density and momentum.

However, our findings emphasize the potential significant risks associated with applying such an informed  NN to our specific problem. This concern is evident in the right panel of Fig.~\ref{fig:error}. 
While the absolute error and the error observed in $N_{{\nu_e} + {\bar\nu_e}}$ are comparable to what one 
observes in the left and middle panels, there is a new type of error which appears in the total  number of neutrinos  and its first moment, soaring to values as high as 1-3\%.
Such  errors, whether in total neutrino number density or momentum, have the potential to  distort the physics of CCSNe and NSMs.
Note that such errors can generate greater risks in scenarios where a disparity might exist between the training and test datasets.

All calculations presented so far  employ a feedforward NN with a single hidden layer containing $n_h = 50$ neurons. The rationale for this choice of the number of neurons is illustrated in Fig.~\ref{fig:num_hid}, where different errors are shown for different NN architectures. It is evident that the NNs perform best on the validation set once $n_h\gtrsim 50$.
It is also illuminating to note that if, for any reason such as computational constraints, a simpler NN with a smaller $n_h$ is used, the performance of the PINN surpasses notably that of the model without additional features. However, this performance gap diminishes as larger $n_h$ values are utilized. Furthermore, it is evident that even the model without explicitly enforced conservation laws achieves its optimal performance when $n_h\gtrsim 50$.

\begin{figure} [tbh!]
\centering
\begin{center}
\includegraphics*[width=.45\textwidth, trim= 10 0 30 20, clip]{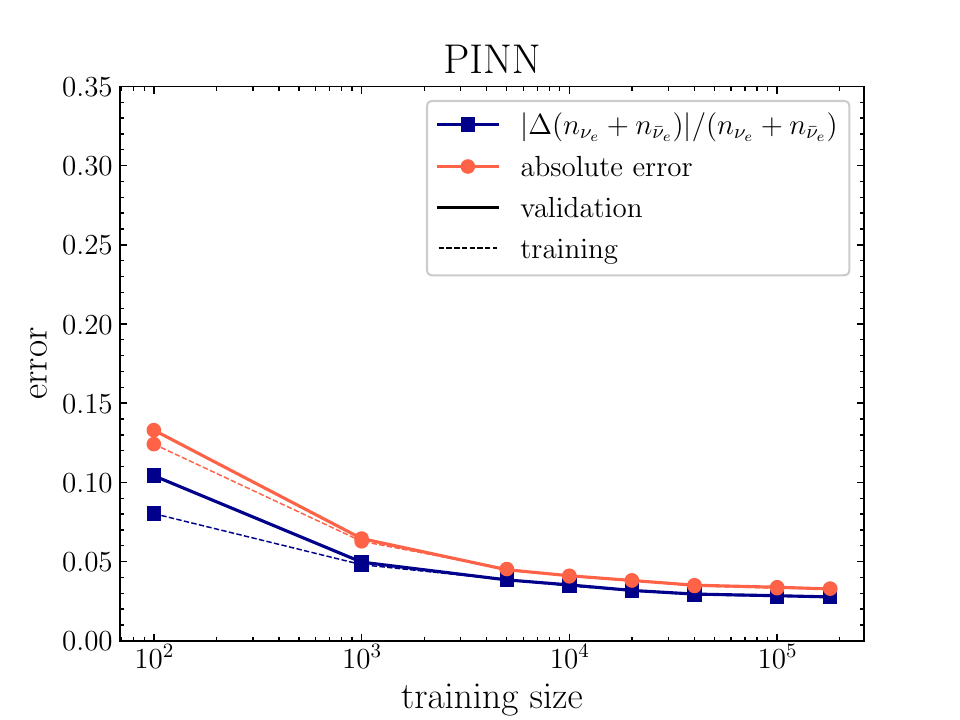}
\end{center}
\caption{
Absolute error in the output of our PINN (red curve) vs the relative error in the number of neutrinos
in electron channel, i.e., $N_{{\nu_e} + {\bar\nu_e}}$ (blue curve).
 Note that having a few thousand
data points in the training set already leads to an error $\lesssim 5\%$, and also the disappearance of error variations between the validation and training sets.}
\label{fig:PINN_error}
\end{figure}

In Fig.~\ref{fig:PINN_error}, we present an analysis of the performance of our PINN as a function of the size of the training set. The red curve represents the absolute error in the PINN's output, while the blue curve illustrates the relative error in $N_{{\nu_e} + {\bar\nu_e}}$.
It is noteworthy that as the training dataset expands to incorporate several thousand data points, the error rapidly diminishes to values  below 5\% and the difference between the error in the validation and training set disappears.
This establishes the absolute minimum number of data points essential for conducting dependable calculations using NN's. However, it's important to bear in mind  that this requisite number is expected to inherently grow as one explores increasingly intricate models involving more inputs and outputs. It is also interesting to  observe that the performance of our NN remains satisfactory even when trained on relatively small datasets, comprising just a few hundred data points.

\subsection{Two-flavor scenario}


\begin{figure*} [tb!]
\centering
\begin{center}
\includegraphics*[width=1.\textwidth, trim= 10 10 0 10, clip]{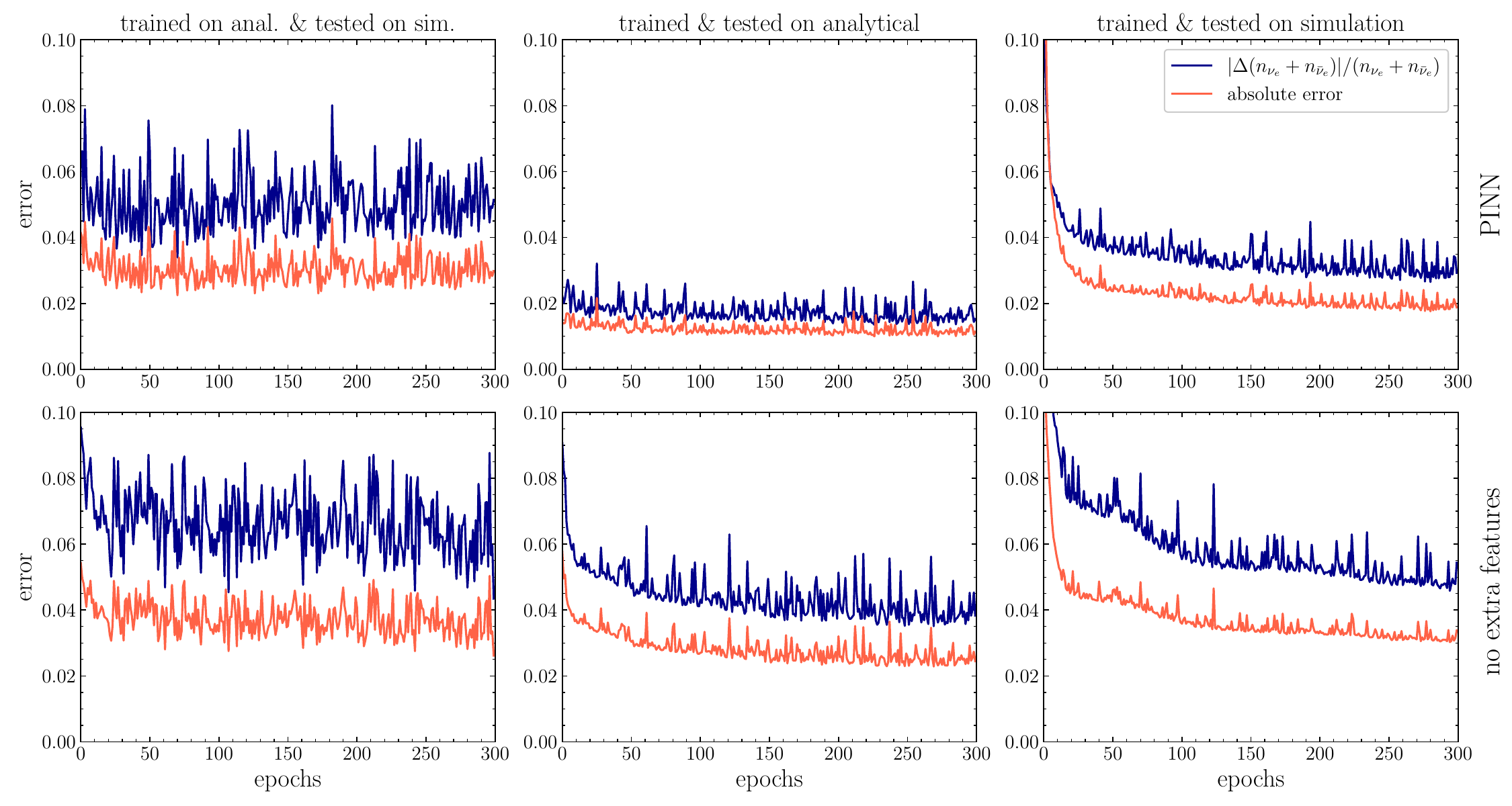}
\end{center}
\caption{
The performance of our NNs in the two-flavor scenario for the PINN and the NN with no extra features.
\sa{The left, middle, and right panels present the performance of the NN model trained using artificial data (with analytical prescription)
and tested on simulation data, the one trained and tested on the artificial data, and the model 
trained and tested on the simulation data, respectively.}
Note that  the PINN continues to outperform especially when it comes to the error in the total number of neutrinos in the electron channel. }
\label{fig:2f_error}
\end{figure*}

In this section, we conduct an evaluation of the performance of our NNs within a two-flavor scenario. The architectural configuration of the NNs and the training process closely mirror those detailed in the preceding section. 
\sa{A notable departure from the prior section lies in the fact that, in this case, we employ the two-flavor version of the survival probability (Eq.~(\ref{eq:sur})). Additionally, we have  considered the results derived from our 1D box simulations of FFCs. These simulations encompass 10,000 \mrw{data} points characterized by the Gaussian angular distributions.
It is crucial to emphasize that when working with the simulation results, no analytical prescription is employed for the survival probability.} Instead, we derive the outcomes of FFCs directly from the simulations. This distinctive approach offers the advantage of enabling us to assess the reliability of the NN models trained on artificial data, when tested on outcomes derived from actual simulations of FFCs.


The left panels of Fig.~\ref{fig:2f_error} present the performance of our NNs trained using the artificial data, when tested on the simulation data. Notably, the PINN consistently outperforms the basic NN with no extra features, particularly in the error associated with $N_{{\nu_e} + {\bar\nu_e}}$. It is crucial to observe that the errors exhibit more pronounced variations compared to previous cases. This increased variability might be attributed to inherent systematic errors in the analytical prescription (for an estimation of the error associated with the analytical prescription, see Table.1 of Ref.~\cite{Xiong:2023vcm}).

During the evaluation of our NNs' performance on simulation data, we noted the critical importance of ensuring the equivalence between the input parameter space covered in the training set and the test set. This holds particularly true for the ranges of $\alpha$ and $\alpha_{\nu_x}$, as well as for  the hierarchical relationship among $F_\nu$'s.
To elaborate further, if the test set encounters regions within the input parameter space that were completely unseen during the training phase, it could significantly degrade the performance of the NN, potentially resulting in very poor performance.

The middle and right panels of Fig.~\ref{fig:2f_error} display errors encountered during the computations in which both training and testing were performed on identical datasets—either artificial data generated using analytical formulas or data obtained from simulations.
  As discussed before, we have considered here  a single form of neutrino angular distributions, namely the Gaussian one. This  analysis provides  insights into the 
inherent errors presented in each of the training sets. Then by comparing them with the left panels, one gets an idea of the error existing in the analytical formula.

The results depicted in Fig.~\ref{fig:2f_error} demonstrate the \sa{notable} enhancement achieved by utilizing our PINN method in the case of two-flavor scenario. Specifically, a noticeable disparity is evident between 
the performance of PINN and that of the NN with no extra features,
surpassing the distinctions observed in Fig.~\ref{fig:error} for the three-flavor scenario. This fundamental discrepancy between two- and three-flavor scenarios highlights a greater degeneracy in the former, ultimately resulting in a more  overall performance improvement when additional information, such as the implementation of PINN, is incorporated.

It is also worth noting a shift in the hierarchy between the absolute error and the error in $N_{{\nu_e} + {\bar\nu_e}}$, in the two- and three-flavor scenarios (compare Fig.~\ref{fig:2f_error} with the left and middle panels of Fig.~\ref{fig:error}). Although this observation is intriguing, it is crucial to recognize that comparing an absolute error with a relative error may not be entirely equitable. Such a hierarchy is anticipated to be sensitive to changes in the data, and thus, its intrinsic merit is limited. To address this concern, we investigated the hierarchy between the \emph{relative} absolute error and the relative error in $N_{{\nu_e} + {\bar\nu_e}}$. This led us to find that the hierarchy remains consistent when considering these two types of errors, providing a fair basis for comparison.

Despite what discussed above, we opted to utilize absolute error instead of relative absolute error throughout this study for two primary reasons. Firstly, in our calculations, we have already normalized all quantities by $n_{\nu_e}$, resulting in having relative values for each quantity. This implies that any absolute error could be interpreted already to be relative in spirit. Secondly, to avoid excessive sensitivity to the small values associated with some of the neutrino quantities, we found it more appropriate to employ absolute error as a metric in our study. This decision ensures a balanced and meaningful evaluation of our results.

It is  important to note that our NNs are not specifically designed to capture $\nu$ELN crossings.
Hence, a practical implementation involves first employing one of the classical ML methods outlined in Refs.~\cite{abbar2023applications,Abbar:2023zkm}. Subsequently, if fast modes are identified, our NNs can be effectively utilized. Despite this, given the broad range of our training set, which includes cases with both narrow and shallow crossings, our method is expected to provide accurate results, even in the absence of a crossing (where our approach should return outputs that are very close to the initial values).

\section{DISCUSSION AND OUTLOOK}\label{sec:dis}

Intensive simulations have demonstrated that FFCs can achieve equilibrium  states in some models. In this study, we have employed neural networks (NNs) to predict the asymptotic outcome of FFCs in a three-flavor neutrino gas within a 1D box with periodic boundary conditions. Specifically, our focus was on the first two moments of neutrino angular distributions as inputs/outputs, \sa{making} our NN models applicable to cutting-edge CCSNe and NSM simulations. 
We have shown that  our NNs  can predict the asymptotic outcomes of the (anti)neutrino $I_0$'s and $I_1$'s with a notable accuracy, corresponding to an error of $\lesssim 3\%$.

In order to enhance the performance of our NNs, we implement some novel features aiming at capturing the  characteristics of the expected neutrino survival probability distributions. 
Firstly, we incorporate a new feature related to the position of the zero crossing in the distribution of $\nu$ELN, $\mu_c$. Additionally, we introduce another feature indicating on which side of $\mu_c$ the expected equipartition between different neutrino flavors  occurs. Both of these  features are derived through a layer of regression applied to the initial inputs of the NN (see Fig.~\ref{fig:scheme}).

In the context of neutrino flavor conversions in CCSNe and NSMs, a critical parameter is the quantity of neutrinos and antineutrinos in the electron channel. To further optimize our NNs, we incorporate a supplementary loss term penalizing any discrepancies in predicting $N_{{\nu_e} + {\bar\nu_e}}$. 
The results demonstrate a relative improvement in our customized physics-informed neural network (PINN) due to the incorporation of extra features and a tailored loss function, outperforming a basic neural network that uses a standard mean squared error loss function and lacks these extra features.

We have also conducted a comprehensive evaluation of the performance of our NNs focusing on the variance between the training and validation sets (Fig.~\ref{fig:PINN_error}). Our findings reveal that the observed variance almost disappears when considering a minimum of a few thousand data points. 
This establishes an absolute minimum number of data points essential for developing a dependable NN for predicting the outcome of FFCs in our model. 

An intriguing observation arising from our study is that, even with the utilization of relatively small datasets, the variance remains modest, with an associated error  limited to $\lesssim 15\%$. This insight further emphasizes the potential of NNs in scenarios where obtaining extensive datasets may be challenging or resource-intensive.

Instead of rigidly adhering to the strict conservation of neutrino quantities, we have also assessed the performance of a NN with a more flexible approach, where all the neutrino moments ($I_0$'s and $I_1$'s for all flavors) are treated as outputs. Here, we introduced an additional loss term effectively ensuring the conservation of (anti)neutrino number densities and momenta.
However, our research has underscored noteworthy concerns associated with the application of such an informed NN to our specific problem. Indeed, we have shown that  there could exist unignorable errors in the total number of neutrinos and their first moments, indicating a capacity to distort the physics of CCSNe and NSMs. 

In our research, our NN models were predominantly trained on artificial data derived from two initial parametric angular distributions: the maximum entropy and Gaussian distributions. Additionally, we assessed the performance of our NN models using simulation data in a two-flavor scenario. Our findings indicate that the observed errors can be rationalized by accounting for the anticipated discrepancies between analytical and numerical results, as well as the inherent errors present in the training set.

In summary, our findings underscore the viability of NNs in forecasting the asymptotic outcomes of FFCs, once only the initial two moments of neutrino angular distributions are taken into account. This marks a significant advancement in the potential integration of FFCs into simulations of CCSNe and NSMs. However, there are still \sa{vital} avenues for further exploration.
Firstly, our NN models were notably constrained to scenarios where neutrino distributions were assumed to be axisymmetric. Additionally, we operated under the assumption that $\nu_x$ and $\bar\nu_x$ exhibit similar distributions. Relaxing these assumptions necessitates access to training datasets derived from actual simulations of FFC evolution in models without imposed axisymmetry, and where $\nu_x$ and $\bar\nu_x$ distributions may differ. Furthermore, our results are founded on a single-energy neutrino gas, prompting a crucial question regarding expectations in a multi-energy neutrino environment. This consideration is especially \sa{relevant}, given that almost all practical applications  involve predicting FFC outcome regarding the neutrino energy spectrum.
Given the 
efficacy of NNs in this domain,
\sa{taking these crucial steps enhances remarkably}
 the feasibility of incorporating FFCs into CCSN and NSM simulations.


\section*{Acknowledgments}
\sa{We are deeply grateful to Georg Raffelt for insightful conversations and reading our manuscript.}
We also thank Gabriel Mart\'inez-Pinedo and Oliver Just for fruitful discussions.
S.A. was supported by the German Research Foundation (DFG) through
the Collaborative Research Centre  ``Neutrinos and Dark Matter in Astro-
and Particle Physics (NDM),'' Grant SFB-1258, and under Germany’s
Excellence Strategy through the Cluster of Excellence ORIGINS
EXC-2094-390783311.
M.-R.~W.\ acknowledges supports from the National Science and Technology Council under Grant No.~111-2628-M-001-003-MY4, the Academia Sinica under Project No.~AS-CDA-109-M11, and the Physics Division, National Center for Theoretical Sciences, as well as the resource of the Academia Sinica Grid-computing Center (ASGC).
Z.X. was supported by the European Research Council (ERC) under the European Union's Horizon 2020 research and innovation programme (ERC Advanced Grant KILONOVA No.~885281).
We would also like to acknowledge the use of the following softwares: \textsc{Scikit-learn}~\cite{pedregosa2011scikit}, 
\textsc{Keras}~\cite{chollet2015keras},
\textsc{Matplotlib}~\cite{Matplotlib}, \textsc{Numpy}~\cite{Numpy}, \textsc{SciPy}~\cite{SciPy}, and \textsc{IPython}~\cite{IPython}.

\bibliographystyle{elsarticle-num}
\bibliography{Biblio}

\clearpage

\end{document}